\documentstyle [12pt] {article}
\begin{document}

\pagestyle{empty}
\rightline{\vbox{
\halign{&#\hfil\cr
&NUHEP-TH-95-08\cr
&UCD-95-23\cr
&hep-ph/9507398\cr
&July 1995\cr}}}
\bigskip
\bigskip
\bigskip
{\Large\bf
	\centerline{Gluon Fragmentation}
	\centerline{into Spin-Triplet S-Wave Quarkonium}}
\bigskip
\normalsize

\centerline{Eric Braaten}
\centerline{\sl Department of Physics and Astronomy, Northwestern University,
    Evanston, IL 60208}
\bigskip

\centerline{Tzu Chiang Yuan}
\centerline{\sl Davis Institute for High Energy Physics}
\centerline{\sl Department of Physics, University of California,
    Davis, CA  95616}
\bigskip

\begin{abstract}
The leading color-singlet contribution to the
fragmentation function for a gluon to split into spin-triplet
S-wave quarkonium is presented.   In the case of charmonium,
we find that this color-singlet term is always negligible compared to the
leading color-octet contribution.
\end{abstract}

\vfill\eject\pagestyle{plain}\setcounter{page}{1}

The dominant mechanism for the production
of heavy quarkonium at sufficiently large transverse momentum
is parton fragmentation \cite{gswave}.
In Ref.~\cite{gswave}, we calculated the fragmentation
functions for gluons to split into S-wave quarkonium
states to leading order in $\alpha_s$
and to leading order in the nonrelativistic expansion.
The fragmentation function for the $^1S_0$ state was given in
analytic form, but only numerical results were
presented for the $^3S_1$ state due to the lengthy and
cumbersome expression for the fragmentation function.
In this Brief Report, we present the detailed formula for the gluon
fragmentation function into the $^3S_1$ state.

A rigorous theory of the inclusive production of heavy quarkonium has been
developed recently by Bodwin, Braaten, and Lepage \cite{bbl}.
This approach is based on the use of nonrelativistic QCD (NRQCD)
to factor the production rate into short-distance
coefficients that can be computed in perturbation theory and long-distance
factors that are expressed as NRQCD matrix elements.
Using this formalism, the fragmentation function for a gluon to
split into a quarkonium state $X$ with longitudinal momentum fraction $z$
can be written as
\begin{equation}
D_{g \rightarrow X} (z, \mu) = \sum_n d_n(z,\mu)
\langle {\cal O}^X_n \rangle \; ,
\label{Dfac}
\end{equation}
where ${\cal O}^X_n$ are local 4-fermion operators  in NRQCD.
The short-distance coefficients $d_n(z,\mu)$ are independent of the quarkonium
state $X$. For a fragmentation scale $\mu$ of order of the heavy quark mass
$m_Q$, they can be computed using perturbative theory in $\alpha_s(2 m_Q)$.
The dependence on the quarkonium state $X$ appears in the long-distance
factors $\langle {\cal O}^X_n \rangle$. The relative
magnitude of the various matrix elements for a given state $X$ can be
estimated by how they scale with $m_Q$ and with the typical relative
velocity $v$ of the heavy quark inside the quarkonium. Thus, the factorization
formula (\ref{Dfac}) for the gluon fragmentation function is a
double expansion in $\alpha_s$ and $v$. To determine the relative importance
of the terms in this formula, one should take into account both the
scaling in $v$ of the matrix elements and the
order in $\alpha_s$ of their coefficients.

We now consider the fragmentation function for a $^3S_1$ state which
we denote by $V$.
In the color-singlet model for quarkonium production \cite{schuler},
only a single term in the expansion (\ref{Dfac}) for the fragmentation function
is retained.  In the notation of Ref.~\cite{bbl},
the matrix element in this term is
$\langle {\cal O}^V_{1}(^3S_1) \rangle$.
This matrix element is proportional to the probability for the formation
of the state $V$ from a point-like $Q \overline Q$ pair in a
color-singlet $^3S_1$
state.  The leading contribution to its short-distance coefficient
$d_1^{(^3S_1)}(z)$ arises from the
parton process $g^* \to Q \overline Q gg$ and is of order $\alpha_s^3$.
Using the velocity-scaling rules of \cite{bbl}, the matrix element
$\langle {\cal O}^V_{1}(^3S_1) \rangle$
is of order $v^3$, so the contribution to the gluon
fragmentation function is of order $\alpha_s^3 v^3$.
The factorization approach reduces to the color-singlet model
in the limit $v \to 0$, since all other matrix elements in the
expansion (\ref{Dfac}) are higher order in $v^2$.

The charmonium and bottomonium systems
are probably not sufficiently nonrelativistic that matrix elements
that are suppressed by powers of $v^2$ can be completely neglected.
In the case of a $^3S_1$ state, there is one matrix element that is
suppressed by a single power of $v^2$ relative to
$\langle {\cal O}^V_1(^3S_1) \rangle$,
but it also has a short-distance
coefficient $d(z)$ of order $\alpha_s^3$.  There are several
``color-octet'' matrix elements
that are suppressed by two powers of $v^2$, including
$\langle {\cal O}^V_8(^1S_0) \rangle$,
$\langle {\cal O}^V_8(^3S_1) \rangle$, and
$\langle {\cal O}^V_8(^3P_J) \rangle$.
Of particular importance in gluon fragmentation is the
matrix element $\langle {\cal O}^V_8(^3S_1) \rangle$,
because it has a short-distance coefficient of order $\alpha_s$
due to the parton process $g^* \to Q \overline Q$.  All other matrix elements
that are suppressed by $v^4$ have short-distance coefficients of order
$\alpha_s^2$ or higher.  The matrix element
$\langle {\cal O}^V_8(^3S_1) \rangle$
is proportional to the probability for the formation
of the state $V$ (plus other particles)
from a point-like $Q \overline Q$ pair in a color-octet $^3S_1$ state.
The corresponding contribution to the
fragmentation function is of order $\alpha_s v^7$, compared
to $\alpha_s^3 v^3$ for the leading color-singlet term.
The two fewer powers of $\alpha_s$ can compensate for the suppression of the
matrix element by $v^4$.  Keeping only these two terms, the
gluon fragmentation function for the $^3S_1$ state at the initial scale
$2 m_Q$ can be written as
\begin{equation}
D_{g \rightarrow V}(z, 2 m_Q) \;=\;
d_1^{(^3S_1)}(z,2 m_Q) \langle {\cal O}^V_1(^3S_1) \rangle
\;+\; d_8^{(^3S_1)}(z, 2 m_Q) \langle {\cal O}^V_8(^3S_1) \rangle \; .
\label{Dfrag}
\end{equation}

To leading order in $\alpha_s$, the coefficient of the
color-singlet matrix element in (\ref{Dfrag}) can be deduced
from the Feynman amplitude for $g^* \rightarrow Q \overline Q g g$,
where the $Q \overline Q$ pair is produced in a color-singlet $^3S_1$ state
with vanishing relative momentum.
The square of the amplitude
for this process can be extracted from a calculation of the matrix element
for $e^+ e^- \rightarrow \psi g g$ \cite{ks}. There are several
typographical errors that must be corrected in eq. (5)
of Ref.~\cite{ks}. In the sixth term on the right side,
the factor $(1 - \mu^2 - x_2^2)$ should be $(1 - \mu^2 - x_2)^2$.
In the last term, $(1 - \mu - x_1)^2$ should be $(1 - \mu^2 - x_1)^2$,
$(1 - \mu - x_2)^2$ should be $(1 - \mu^2 - x_2)^2$,
$(K_- \cdot L_-)^2$ should be $(K_- \cdot L_-)$, and the overall
sign of this last term should be changed from minus to plus.
Having made these corrections, one can reproduce previous results
for the energy distribution for $\gamma^* \to \psi g g$ \cite{keung}.
The fragmentation function for the $^3S_1$ state can be
calculated  by the same method used for  the $^1S_0$
state in Ref.~\cite{gswave}.
The calculation is rather involved and we present only the final result:
\begin{eqnarray}
d_1^{(^3S_1)}(z, 2 m_Q)
\;=\; {5 \over 5184 \pi m_Q^3} \alpha_s(2 m_Q)^3
\int_0^z dr \int_{(r+z^2)/2z}^{(1+r)/2} dy \;
{1 \over (1-y)^2 (y-r)^2 (y^2-r)^2}
\nonumber \\
\times \sum_{i=0}^2 z^i \left( f_i(r,y) \;+\; g_i(r,y)
	{1+r-2y \over 2 (y-r) \sqrt{y^2-r}}
	\log{y-r + \sqrt{y^2-r} \over y-r - \sqrt{y^2-r}} \right) \;.
\label{Dpsi} \end{eqnarray}
The integration variables are $r = 4 m_Q^2/s$ and  $y = p \cdot q/s$,
where $p$ and $q$ are the 4-momenta of the quarkonium and the fragmenting
gluon and $s = q^2$.
The functions $f_i$ and $g_i$ are
\begin{eqnarray}
f_0(r,y) &=& r^2(1+r)(3+12r+13r^2) \;-\; 16r^2(1+r)(1+3r)y
\nonumber \\
&-& 2r(3-9r-21r^2+7r^3)y^2
\;+\; 8r(4+3r+3r^2)y^3 \;-\; 4r(9-3r-4r^2)y^4
\nonumber \\
&-&  16(1+3r+3r^2)y^5 \;+\; 8(6+7r)y^6 \;-\; 32 y^7 \;,
\label{f0} \\
f_1(r,y) &=& -2r(1+5r+19r^2+7r^3)y \;+\; 96r^2(1+r)y^2
\;+\; 8(1-5r-22r^2-2r^3)y^3
\nonumber \\
&+& 16r(7+3r)y^4 \;-\; 8(5+7r)y^5 \;+\; 32y^6 \;,
\label{f1} \\
f_2(r,y) &=& r(1+5r+19r^2+7r^3) \;-\; 48r^2(1+r)y \;-\; 4(1-5r-22r^2-2r^3)y^2
\nonumber \\
&-& 8r(7+3r)y^3 \;+\; 4(5+7r)y^4 \;-\; 16y^5 \;,
\label{f2} \\
g_0(r,y) &=& r^3(1-r)(3+24r+13r^2) \;-\; 4r^3(7-3r-12r^2)y
\;-\; 2r^3(17+22r-7r^2)y^2
\nonumber \\
&+& 4r^2(13+5r-6r^2)y^3 \;-\; 8r(1+2r+5r^2+2r^3)y^4
\;-\; 8r(3-11r-6r^2)y^5
\nonumber \\
&+& 8(1-2r-5r^2)y^6 \;,
\label{g0} \\
g_1(r,y) &=& -2r^2(1+r)(1-r)(1+7r)y \;+\; 8r^2(1+3r)(1-4r)y^2
\nonumber \\
&+& 4r(1+10r+57r^2+4r^3)y^3
\,-\, 8r(1+29r+6r^2)y^4 \,-\, 8(1-8r-5r^2)y^5 ,
\label{g1} \\
g_2(r,y) &=& r^2(1+r)(1-r)(1+7r) \;-\; 4r^2(1+3r)(1-4r)y
\nonumber \\
&-& 2r(1+10r+57r^2+4r^3)y^2
\,+\, 4r(1+29r+6r^2)y^3 \,+\, 4(1-8r-5r^2)y^4.
\label{g2} \end{eqnarray}
The integrals over $r$ and $y$ in (\ref{Dpsi}) must be evaluated numerically
to obtain the fragmentation function at the energy scale $\mu = 2 m_Q$.
The fragmentation function at higher energy scales $\mu$
is then obtained by Altarelli-Parisi evolution.

The short-distance coefficient of the color-octet matrix element
in (\ref{Dfrag}) can be calculated to leading order in $\alpha_s$
from the Feynman amplitude for $g^* \rightarrow Q \overline Q$.
The result is \cite{gpwave}
\begin{equation}
d_8^{(^3S_1)}(z,2 m_Q) = \frac{\pi \alpha_s(2 m_Q)}{24 m_Q^3} \delta (1-z)
\; .
\end{equation}
The radiative correction of order $\alpha_s^2$ has also been calculated
recently by Ma \cite{ma}.

The relative importance of the color-singlet and color-octet
contributions to the fragmentation functions can be determined
by integrating the initial fragmentation function (\ref{Dfrag})
over $z$ to get the fragmentation probability at the scale $2 m_Q$  :
\begin{eqnarray}
\int_0^1 dz \; D_{g \rightarrow V}(z, 2 m_Q) &=&
(8.28 \times 10^{-4}) {\alpha_2^3(2 m_Q) \over m_Q^3}
	\langle {\cal O}^V_1(^3S_1) \rangle
\nonumber \\
&& \;+\; (1.31 \times 10^{-1}) {\alpha_s(2 m_Q) \over m_Q^3}
	\langle {\cal O}^V_8(^3S_1) \rangle \; .
\label{Pfrag}
\end{eqnarray}
The value of the color-singlet matrix element
$\langle {\cal O}^V_{1}(^3S_1) \rangle$
can be determined from the electronic width of the vector meson state.
In the case of the charmonium states $J/\psi$ and $\psi'$, the
matrix elements $\langle {\cal O}^V_1(^3S_1) \rangle$
are approximately $0.73 \; {\rm GeV}^3$ and $0.11 \; {\rm GeV}^3$,
respectively.
The most reliable determinations of the color-octet matrix elements
come from recent data on prompt charmonium production from the CDF detector
at the Tevatron.
The color-octet contributions are necessary to explain the
magnitude of the cross section for prompt $J/\psi$ and $\psi'$ production
at large transverse momentum,
and they also explain the shape of the transverse
momentum distribution \cite{psiprime}.
The values of the matrix elements $\langle {\cal O}^V_8(^3S_1) \rangle$
that are obtained by fitting
the observed cross sections are approximately
$1.5 \times 10^{-2} \; {\rm GeV}^3$ for $J/\psi$ and
$4.3 \times 10^{-3} \; {\rm GeV}^3$
for $\psi'$ \cite{manganocho}.
Using the value $\alpha_s(2 m_c)= 0.26$, we find that the two
terms in the fragmentation probability (\ref{Pfrag}) are approximately
$3.2 \times 10^{-6}$ and $1.5 \times 10^{-4}$ in the case of $J/\psi$
and $4.7 \times 10^{-7}$ and $4.3 \times 10^{-5}$ in the case of $\psi'$.
The color-octet term is larger by about a factor of 50 for the $J/\psi$
and larger by about a factor of 100 for the $\psi'$.
We conclude that the color-singlet term is always negligible compared to
the color-octet term for charmonium.

In the case of bottomonium, the color-singlet term in the gluon
fragmentation function is probably also negligible.
Because of the running of the coupling constant between the scales
$2 m_c$ and $2 m_b$, the suppression of the color-singlet term by $\alpha_s^2$
decreases its relative importance by about a factor of 2. On the other hand,
the suppression of the color-octet term by $v^4$
decreases its relative importance by about a factor of 10.
The net effect is that the color-singlet term is still likely to
be more than an order of magnitude smaller than the color-octet term.

This work was supported in part by the U.S. Department of Energy,
Division of High Energy Physics, under Grant DE-FG02-91-ER40684 and
Grant DE-FG03-91ER40674.

\vfill\eject

\vfill\eject

\end{document}